\documentstyle[11pt,newpasp,twoside,epsf]{article}
\def\ntot{235} 
\def\ncon{180} 
\def\ncan{55} 
\def\nnhi{75}  
\def\twop{38}  
\def\two{18}  
\def\one{19}  

\def\edcomment#1{\iffalse\marginpar{\raggedright\sl#1\/}\else\relax\fi}
\reversemarginpar

\begin{document}
\title{Metallicity Evolution of Damped Lyman-$\alpha$ Systems}
\author{Sandra Savaglio}
\affil{Space Telescope Science Institute, 3700 San Martin Drive, 
Baltimore MD 21218, USA}
\affil{}

\begin{abstract}
According to Pei, Fall \& Hauser (1999), the global metallicity
evolution of the Universe can be represented by the ratio of the total
metal content to the total gas content measured in Damped
Lyman--$\alpha$ (DLA) systems (the ``column density weighted metallicity''
\`a la Pettini). To minimize dust obscuration effects, a DLA sample
with negligible dust content is considered, namely, 50 DLAs with $\log
N_{HI} < 20.8$. The global metallicity found shows clear evidence of
redshift evolution that goes from $\sim1/30$ solar at $z\sim4.1$ to
solar at $z\sim0.4$. More generally, DLAs with measured heavy elements
probe the ISM of high redshift galaxies. The whole sample collected
from the literature contains \nnhi~DLAs. The metallicity is calculated
adopting for the dust correction the most general method used so far,
based on models of the ISM dust depletions in the Galaxy.  The
intrinsic metallicity evolution of DLA galaxies is $d\log Z_{DLA}/dz
\propto -0.33 \pm 0.06$.

\end{abstract}

\section{Introduction}

A Damped Lyman--$\alpha$ (DLA)  is often referred to as a QSO
absorption line system with HI column density larger than
$2\times10^{20}$ atoms cm$^{-2}$ associated with clouds in
protogalactic disks (Wolfe et al.~1986). In a more modern fashion,
DLAs are mostly neutral gas clouds (HI column density larger than
$\sim10^{19}$ atoms cm$^{-2}$) associated with the interstellar medium
(ISM) of a not defined class of galaxies detected from redshift $z=0$
up to $z=4.6$. Among others, examples of the variety of
morphologies of the emitting counterpart found nearby DLAs are given
by Le Brun et al.~(1997) and Rao \& Turnshek (1998).

The large investment of telescope time dedicated to QSO surveys (Fan
et al.~1999) and specifically to DLA surveys (Rao \& Turnshek 1999)
has brought the discovery of numerous DLAs. We know \ncon~ objects
(Fig.~1) and \ncan~ candidates for a total of \ntot~DLAs in the
redshift range $0.0 < z < 4.6$. Among these, \nnhi~are DLAs for which
the column density of HI and one or more of the following ions have
been measured: FeII, SiII, NiII, MnII, CrII, ZnII.  These ions are the
dominant contributors to the abundances of the corresponding elements
in HI clouds with high column densities, because they all have
ionization potentials below 13.6~eV.  For this reason it is easy and
straightforward to determine the relative heavy element abundance,
just using the standard relation [X/Y] = $\log (N_X/N_Y)_{DLA} - \log
(X/Y)_\odot$.  An important contribution to the richness of this
sample has been provided by Prochaska \& Wolfe (1999), Pettini et
al.~(1997), and Lu et al.~(1996). The heavy element enrichment of DLAs has
been studied recently in detail by Pettini et al.~(1999) and Vladilo
et al. (2000) who considered the ZnII absorption as a tracer of
metallicity, and by Prochaska \& Wolfe (2000), who preferred FeII
measurements instead. Here we try to show that by combining this
information together to build a larger database (\nnhi~DLAs at
$0.0<z<4.4$) and making some assumptions about biases due to dust
obscuration, we can have good evidence that the metallicity of the
Universe as given by DLAs evolves with cosmic time.

\begin{figure}
\plotone{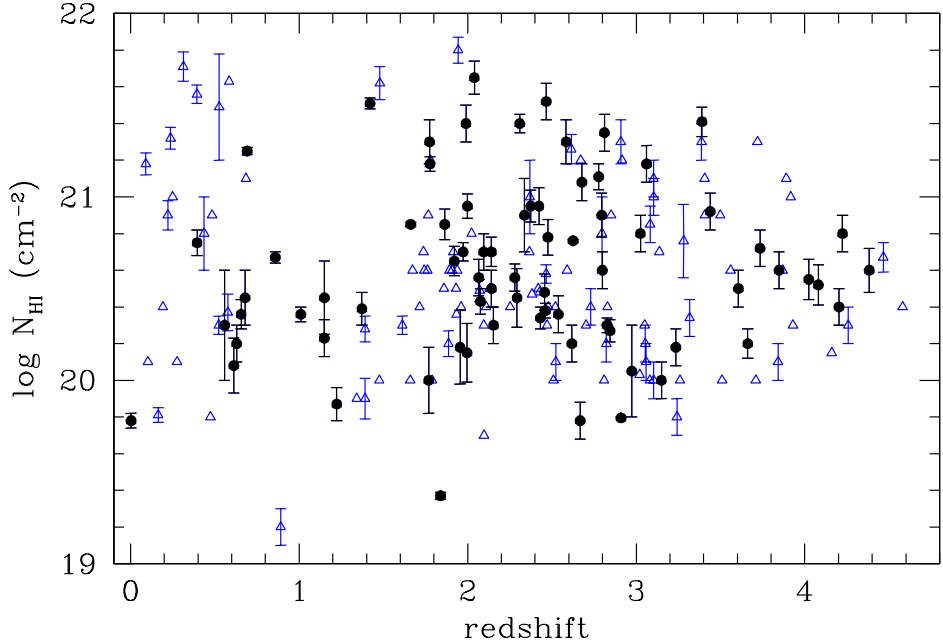}
\caption{Neutral hydrogen column density for a sample of \ncon~DLAs
in the redshift range $0.0<z<4.6$.  Dots are DLAs with at least
one measured heavy element, triangles are the remaining DLAs. Those
points without error bars are DLAs with no HI uncertainty measured.}
\end{figure}

\section{Zinc vs. Iron}

Very often when discussing metallicity in DLAs one is actually talking
about the abundance of zinc relative to hydrogen.  The zinc abundance
is considered the right quantity because it suffers very little
depletion onto dust grains and because the ZnII doublet is never
saturated, and therefore easier to measure with accuracy (Pettini et
al.~1994).  Other elements like iron are often discarded because they are
heavily locked in dust grains.

However important pieces of information are also carried by dust
depleted elements.  In Fig.~2 we plot an example of typical equivalent
widths in the linear part of the curve of growth (Spitzer 1978) of the
ZnII doublet and the multiple lines of FeII as a function of the
transition wavelengths. These are typical of a DLA with $\log
N_{HI}=20.7$ and metallicity 1/10 solar. The presence of dust that
makes the iron in the gas phase less abundant relative to zinc, is
taken into account assuming a relative dust depletion of [Fe/Zn]
$=-0.54$. This value is the mean of iron relative to zinc abundance
measured in a sample of 27 DLAs.

\begin{figure}
\plotone{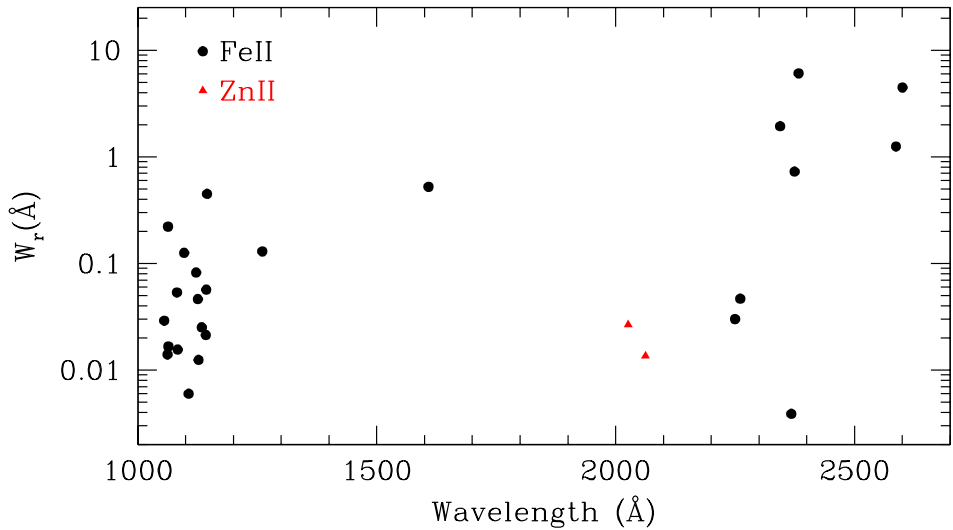}
\caption{Equivalent widths of FeII and ZnII absorption lines for
a DLA with $\log N_{HI} = 20.7$ and metallicity 1/10 solar, as a
function of transition wavelengths. In this case the relative dust depletion 
for the two ions is assumed to be [Fe/Zn] $=-0.54$ dex.}
\end{figure}

The FeII ion is characterized by numerous multiplets of lines which
absorb in a large wavelength range, while ZnII is just present through
a doublet around the 2000 \AA~ rest frame. This means that iron can be
detected in the optical at the lowest and highest redshifts up to
$z\sim 8$, whereas ZnII already approaches the infrared,
where spectroscopy is much less sensitive at $z\sim3.9$.
Another advantage of iron are the oscillator strengths that vary by
three orders of magnitude in different absorption lines, making the
detection of the ion very sensitive in very low metallicity DLAs or overcoming the
problem of saturation when metallicity is high. Even if the ZnII
doublet oscillator strengths are large, the much lower cosmic
abundances of zinc with respect to iron (almost three orders of
magnitudes) make the detection harder when the metallicity is low.
FeII in general is much easier to detect because it is less affected by
biases. This is shown in Fig.~3, where samples of DLAs with
measured FeII and ZnII are reported. In the sample of \nnhi~DLAs there
are 59 objects with FeII in the redshift range $0.0 < z < 4.4$, and 31
objects (almost a factor of two less!)  with ZnII in a smaller
redshift range $0.4 < z < 3.4$. As a consequence if only ZnII is
considered, the statistics are more limited in redshift range and biased
toward low metallicity/dust free DLAs. The selection effects are
discussed in more detail in section 5.

\begin{figure}
\plotone{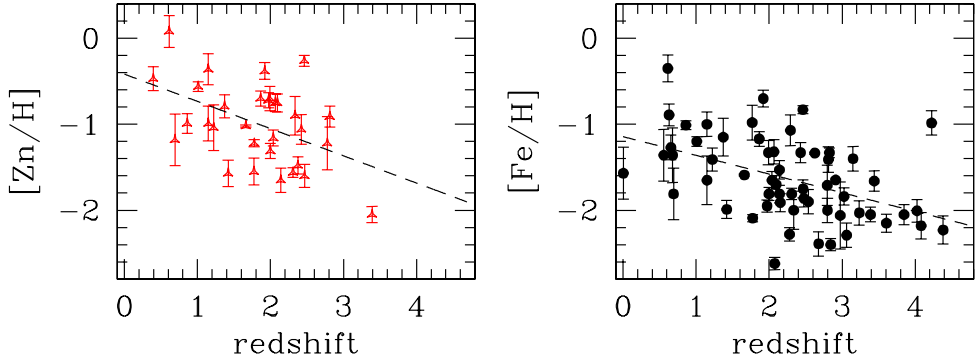}
\caption{Zinc and iron metallicities as a function of redshift in
DLAs. In the first case there are 31 detections at $0.4 < z < 3.4$,
while in the second we have 59 at $0.0 < z < 4.4$. Straight lines are the
linear correlations for the two cases: [Zn/H] $\propto
(-0.32\pm0.11)\times z$ and [Fe/H] $\propto (-0.22\pm0.06)\times z$.}
\end{figure}

\section{Dust Depletion Correction and Metallicity Determination}

To derive a complete picture of the DLA chemical state, one must
consider all detected heavy elements and correct for dust depletion
effects.  This is not particularly easy since every element is
affected differently by dust depletion.  In the Milky Way, a number of
depletion patterns have been identified, showing highest depletions in
dense disk clouds and lowest depletions in low density, warm halo
clouds (Savage \& Sembach 1996).  Although each of the various
patterns shows a range of possible depletions (typically within
factors of 3), the trends are quite clear and the differences between
the various patterns are unambiguously determined. The situation for
DLAs may not be clear a priori because one does not know how similar
the physical conditions are in the absorbing clouds of different
objects nor whether the nature of dust in different objects is
comparable. On the other hand, the similarities of the mean abundance
ratios in DLAs with the ones of warm halo clouds and SMC absorbers
(Welty et al.~1997; Savaglio, Panagia \& Stiavelli~2000) suggest
that the prevailing conditions cannot be very dissimilar. Therefore,
we can make a simplification assuming that the depletion patterns in
DLAs may be reproduced by one of the four depletion patterns
identified for the Milky Way: Warm Halo (WH), Warm Halo + Disk (WHD), Warm
Disk (WD) and Cool Disk (CD) clouds (Savage \& Sembach 1996), thus
modifying the dust--to--metals ratio to obtain the best match to
observations.

In practice, let us call $J$ the depletion pattern for which we are comparing
the abundance ratios observed in a given DLA. For every element
X$_i$ of that DLA, we consider the two quantities:

\begin{equation}
\left\{\begin{array}{ll}
\delta {\rm x}_i = {\rm [X/ H]}_{DLA} -\log \left(Z_{DLA} \over Z_\odot \right) \\
\\
\delta {\rm y}_i = \log \left[1+\frac{k_{DLA}}{k_J}(10^{\delta {\rm x}^J_i}-1)\right]
\end{array}
\right. 
\end{equation}

\noindent 
where $i=1,2,3...$ for S, Zn, Si..., and $J=1,2,3,4$ for WH, WDH, WD,
CD, $\delta {\rm x}^{J}_i$ is the observed depletion of element X$_i$ in
$J$--type clouds.  The two unknowns are the DLA metallicity compared
with solar, $\zeta=Z_{DLA}/Z_\odot$, and its dust--to--metals ratio
$\kappa$ as compared with that of the $J$--type clouds:
$\kappa=k_{DLA}/k_{J}$.  The values of $\zeta$ and $\kappa$ that
minimize the reduced $\chi^2$:

\begin{equation}
\frac{\chi^2(\kappa,\zeta)}{dof} = \frac{1}{N-2} \sum_i \left[\frac{\delta {\rm x}_i-\delta {\rm y}_i}{\sigma (\log {\rm X}_i)}\right]^2
\end{equation} 

\noindent
give the best solution to the problem.  Here $N-2$ is the number of degrees of freedom $dof$ and $\sigma
(\log {\rm X}_i)$ is the error on the measured column density $\log
{\rm X}_i$).  We have applied
this method using the WH, WHD, WD and CD depletion patterns, and for
each DLA we obtained four sets of $\kappa$ and $\zeta$ values, from
which we select the maximum likelihood solution.  As an example we
report in Fig.~4 the depletion pattern of the $z_{DLA}=1.7688$ absorber along the QSO 0216+0803 sight line (Lu et al.~1996) together with the
best solution.

\begin{figure}
\plotone{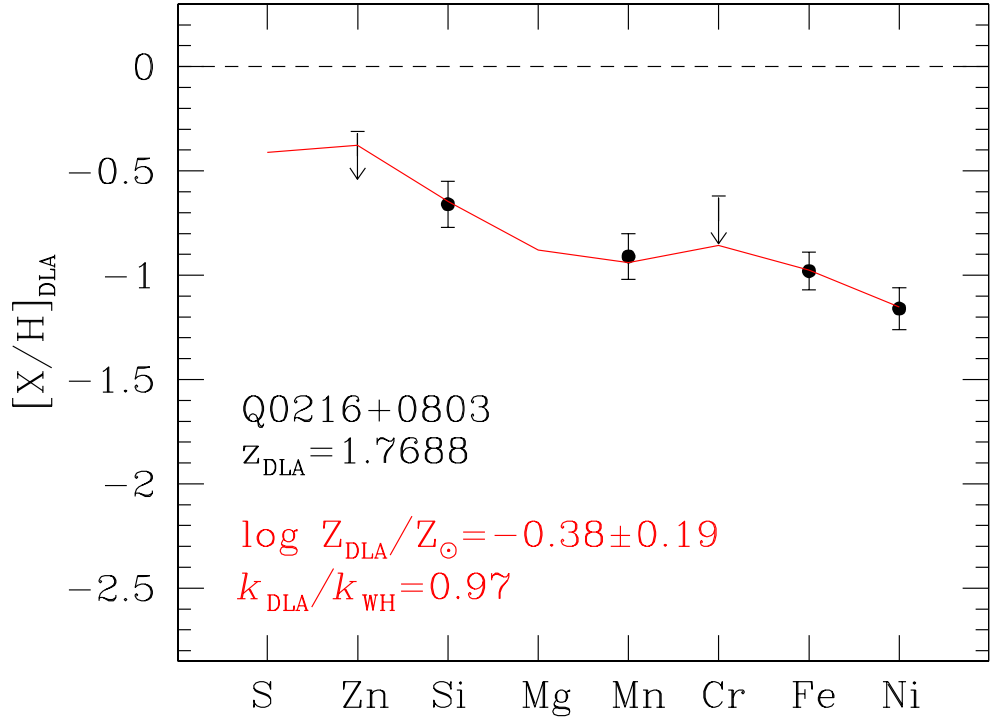}
\caption{Metallicity measurement for the DLA at redshift 1.7688 along
the QSO 0216+0803 sight line. The maximum likelihood solution gives in
this case a metallicity of about 2/5 of solar and a WH depletion
patter with dust--to--metals ratio very similar to that of the WH
clouds. }
\end{figure}

\section{Metallicity Evolution}
 
\begin{figure}
\epsfxsize=.4\hsize
\plotone{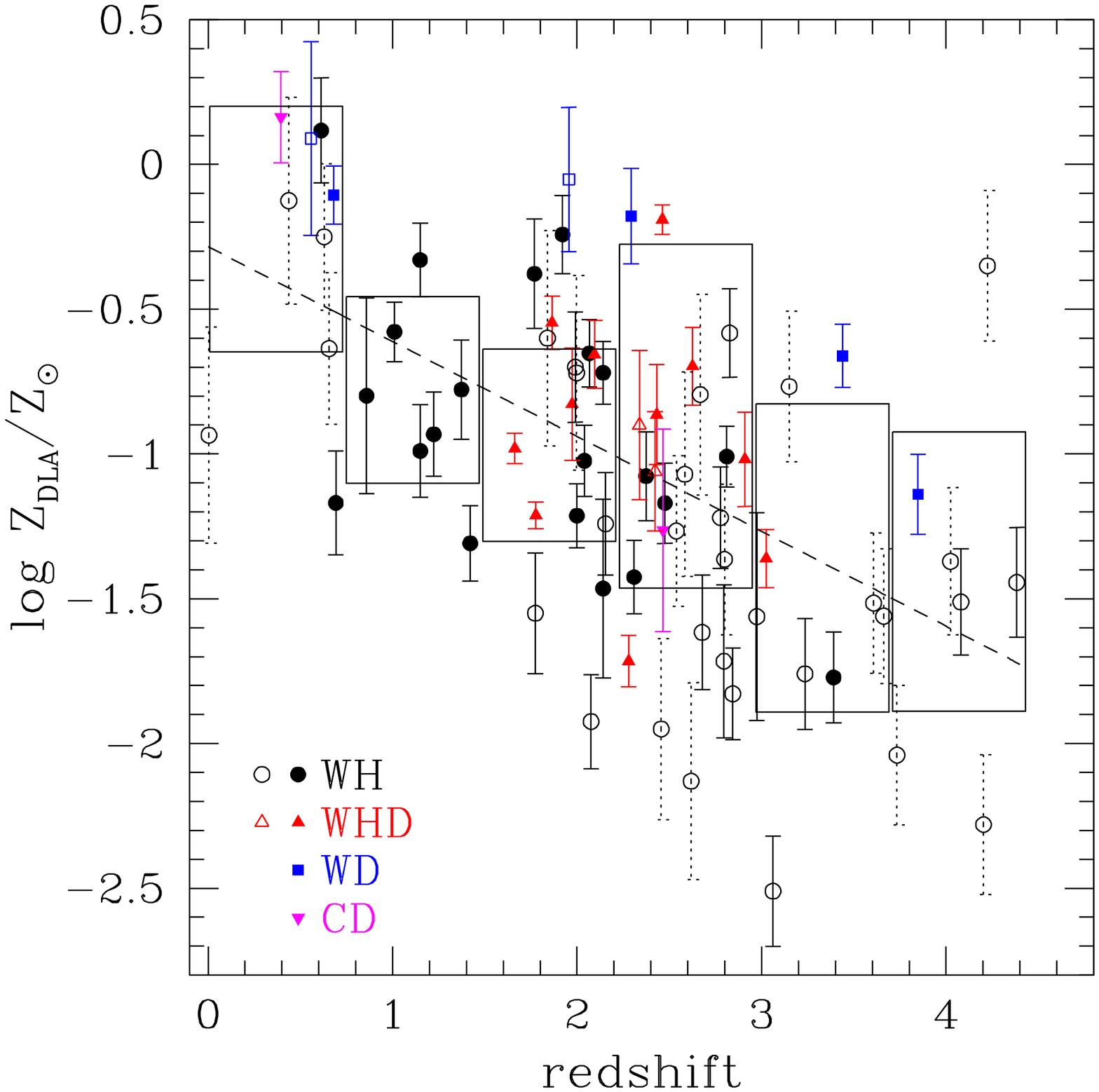}
\caption{Total (gas+dust) metallicity compared to solar for
\nnhi~DLAs. For \twop~objects (filled symbols) the dust correction and
metallicity measurements are the result of a full $\chi^2$
minimization.  Open symbols are DLAs with two elements (solid bars,
\two~objects) or one element (dotted bars, \one~objects) measured.
The dashed line is the best fit for the whole sample in the case of
linear correlation: $d\log Z_{DLA}/dz \simeq -0.33$. Boxes
are centered on the weighted averages over $\Delta z=0.74$ intervals
and vertical widths mark the corresponding $\pm1\sigma$ weighted
dispersion.}
\end{figure}

Fig.~5 shows the metallicity as a function of redshift
found using this method for the \nnhi~DLAs.  Filled symbols represent
DLAs with more than two measured element abundances for which it has
been possible to obtain a proper best fit solution (\twop~DLAs).  For
these, the error bars have been calculated including both the
uncertainties in the measured column densities, and the dust model
fitting errors.  For the cases with only two elements observed, a
reduced $\chi^2$ cannot be calculated; thus, the best fit is
considered less significant (\two~DLAs, empty symbols and solid error
bars). In these cases the error bars include uncertainty in the column
densities plus, summed in quadrature, an average error of 0.15 dex,
selected for being the mean fitting uncertainty for the DLAs with
more than two measured elements.  Finally, for the cases where only
one element is measured, we only tentatively give an estimate of the
metallicity assuming a WH depletion pattern (this is the most likely
case, given the global properties) with $\kappa=1$ (\one~DLAs, empty
symbols and dotted error bars). In these cases the error bars include
an additional $1.5\times0.15$ dex that takes into account the fitting
uncertainty, and this is larger than derived for 92\% of the DLA
sample with more than two measured elements.

The mean metallicity calculated for the whole sample is
$<Z_{DLA}/Z_\odot>\simeq 0.1$ and the mean redshift
$z_{DLA}\simeq2.27$.  In Fig.~5 we also show six boxes, centered on
the weighted averages over $\Delta z=0.74$ intervals.  Their
vertical widths mark the corresponding $\pm1\sigma$ weighted
dispersion.  We note that the vertical dispersion of points of 0.49
dex greatly exceeds the mean error of 0.20 dex, indicating that,
although DLAs follow an average trend, there are real differences
among individual objects in either initial conditions, or time of
formation, or metal enrichment efficiency, or all of the above.  To
see whether there is a redshift gradient of metallicity, we applied
the Spearman test to the whole sample and find a correlation
coefficient of $-0.57$ (99.99\% significance level). The combination
of the largest sample available (\nnhi~DLAs), a large redshift
baseline ($0.0<z<4.4$) and a more accurate dust correction applied
have led to the unambiguous detection of the redshift evolution of
metallicity in DLA galaxies, with mean values around 1/25 of solar at
$z\sim4.1$ to 3/5 of solar at $z\sim0.5$.  If a linear
correlation is assumed between metallicity and redshift, we find a
slope of $d\log Z_{DLA}/dz=-0.328\pm0.057$. To test if this result is
robust, we have calculated the linear correlation for the sample of
\twop~DLAs with more than two measured elements and found consistent
results ($d\log Z_{DLA}/dz = -0.266\pm0.085$, 99.65\% significance
level).

\section{Dust Extinction in DLA QSOs}

\begin{figure}
\epsfxsize=.4\hsize
\plotone{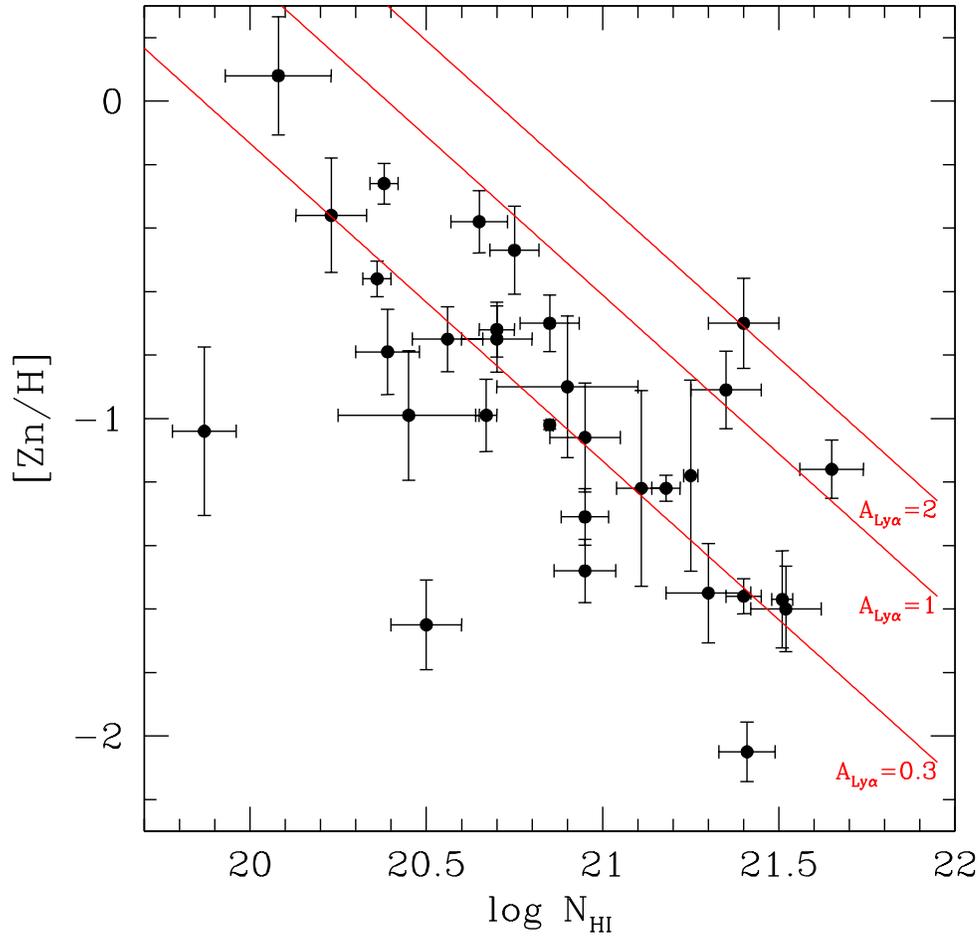}
\caption{Zn abundance compared with solar as a function of HI column
density in 31 DLAs. Straight lines are the [Zn/H] and $\log N_{HI}$
values that give an extinction at Ly$\alpha$ of 0.3, 1 and 2
magnitudes from the bottom to the top (SMC like extinction is
assumed). These correspond to ZnII column densities of $\log N_{ZnII}
= 12.5, 13.0, 13.3$ respectively.}
\end{figure}

In Fig.~6 we show a plot of [Zn/H] vs. $\log N_{HI}$ -- originally
presented by Boiss\'e et al. (1998) -- for a sample of 31 DLAs.  DLAs
with high [Zn/H] abundances and high HI column densities are not
detected.  These are the DLAs with larger dust content and therefore
harder to observe due to dust extinction.  In fact, obscuration at
Lyman--$\alpha$, calculated assuming an SMC--like extinction law
directly proportional to the column density of metals, is
$A_{Ly\alpha}>1$ mag for a ZnII column density $\log N_{ZnII}>13$.  In
other words, dust obscuration can play an important role, in
particular when considering ZnII selected DLAs because ZnII absorption
is harder to detect than that of FeII (see Fig.~2).

The effect of dust obscuration is also shown in Fig.~7 where the
metallicity of high $N_{HI}$ DLAs is displayed
more clearly as a function of redshift .  High $N_{HI}$ DLAs are detected only when the metallicity is
low. The horizontal lines indicate three metallicities and HI column
densities necessary to have an extinction $A_{Ly\alpha}=1$ mag. There
are no DLAs with $\log N_{HI}>21$ at low redshifts. This means that if
there is a metallicity evolution of DLAs, it is difficult to detect in
a sample of high $N_{HI}$ DLAs because dust content, proportional to
metallicity, and thus dust obscuration is larger.

\begin{figure}
\epsfxsize=.4\hsize
\plotone{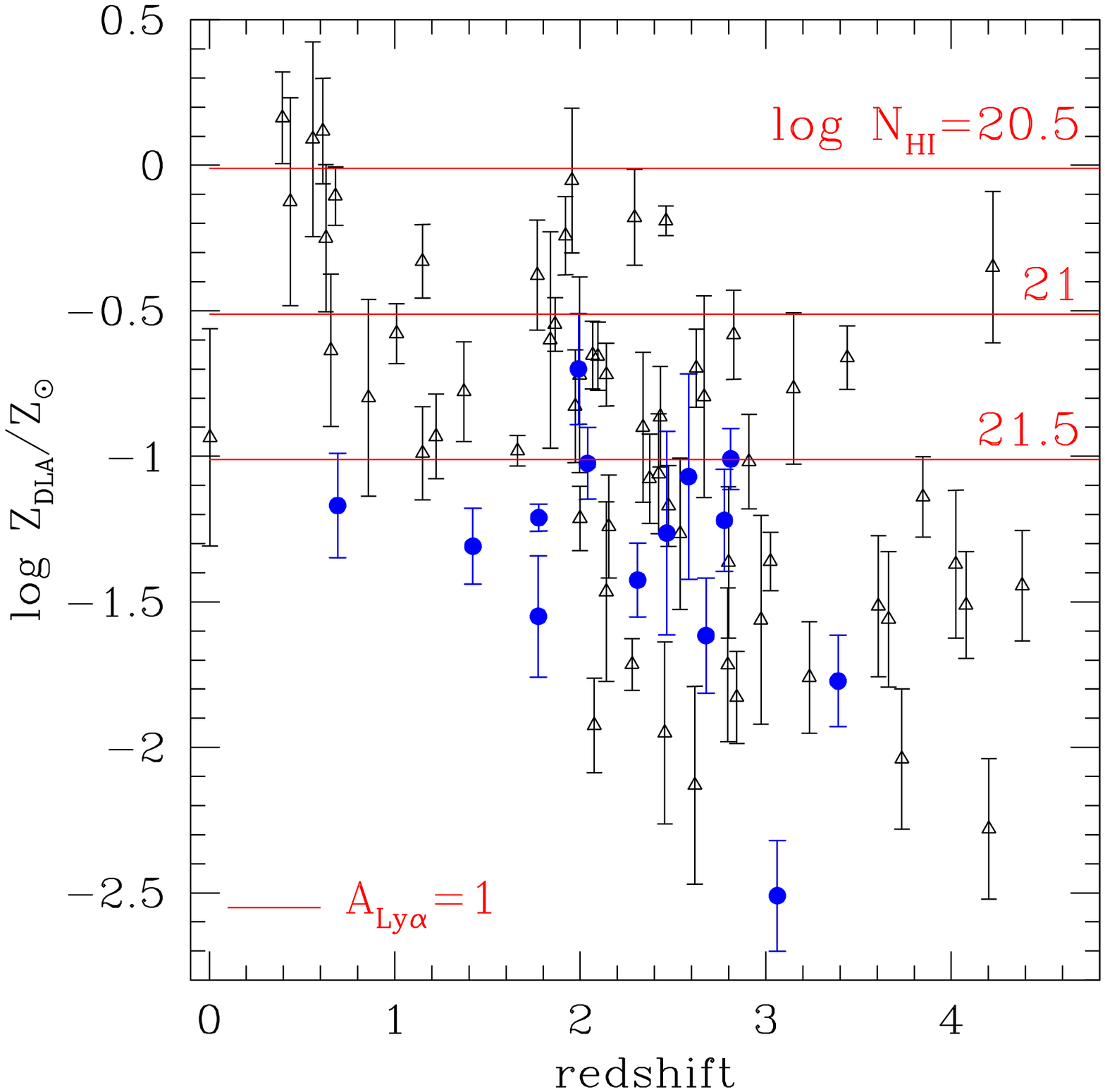}
\caption{Same as Fig.~5, but triangles and dots are DLAs with $\log
N_{HI}\leq21$ and $>21$ respectively.  Horizontal lines indicate a 1
mag extinction at the Ly$\alpha$ wavelength for three different HI
column densities and metallicities. A SMC like extinction curve is
assumed.}
\end{figure}

To better understand a possible dependence on the DLA gas content, we
applied the metallicity analysis to the two subsamples of DLAs with
low ($\log N_{HI} \leq 20.3$, 20 DLAs) and high ($\log N_{HI} \geq
20.9$, 20 DLAs) HI column densities (Fig.~8). What can clearly be
noticed is a completely different redshift distribution in the two
samples due to observational effects. The low $N_{HI}$ DLAs are more
homogeneously distributed in the redshift range than the high $N_{HI}$ ones.
The analysis for the redshift evolution gives a linear correlation
coefficient much more significant for the low $N_{HI}$ sample ($-0.64$,
99.8\% significance level) than for the high $N_{HI}$ sample ($-0.22$, 65.0\%
significance level), while the linear correlation gives a slope of
$d\log Z_{DLA}/dz=-0.40\pm0.11$ and $-0.14\pm0.15$ for the two samples,
respectively.  The fact that the metallicity for high $N_{HI}$ DLAs is
consistent with no evolution appears to be due to the high
concentration of points in the central redshift range (70\% are in the
bin $2.0 < z < 3.1$).  Indeed, the metallicity may evolve just as
rapidly as in the low $N_{HI}$ clouds, as the two slopes differ
only at the $\sim2\sigma$ (95\%) level. This has important
consequences when considering the redshift evolution of the ratio of
the total metal content to the total gas content, as we discuss in the
next section.

\begin{figure}
\plotone{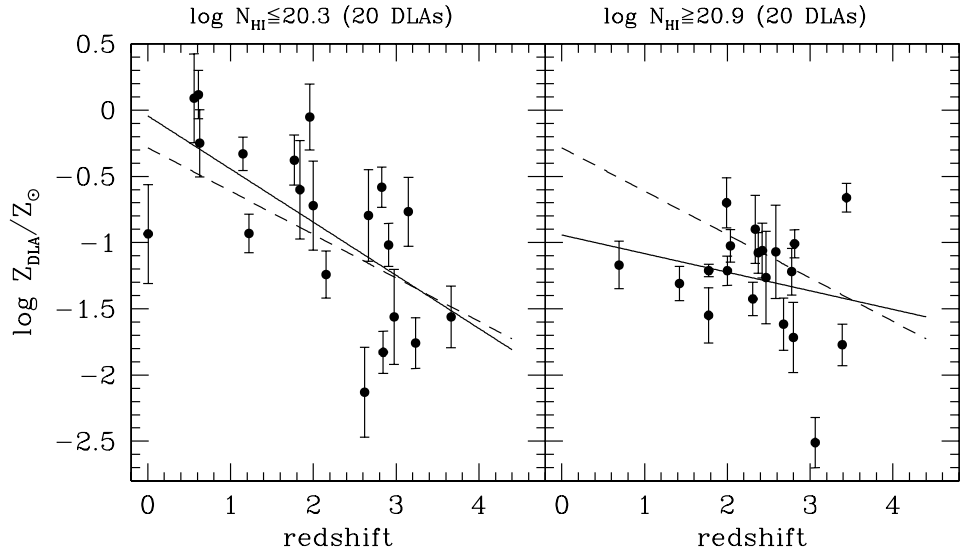}
\caption{Same as Fig.~5, but for two subclasses of DLAs. On the left and
right panels are 20 DLAs with $\log N_{HI}\leq 20.3$ and $\log
N_{HI}\geq 20.9$ respectively. Solid lines are the linear correlation
with $d\log Z_{DLA}/dz = -0.40\pm0.11$ (left panel) and $d\log Z_{DLA}/dz =
-0.14\pm0.15$ (right panel). The dashed line in both plots is the
linear correlation found in the whole sample of Fig.~5: $d\log
Z_{DLA}/dz =-0.328\pm0.057$.}
\end{figure}

\section{Global Metallicity Evolution}

\begin{figure}
\plottwo{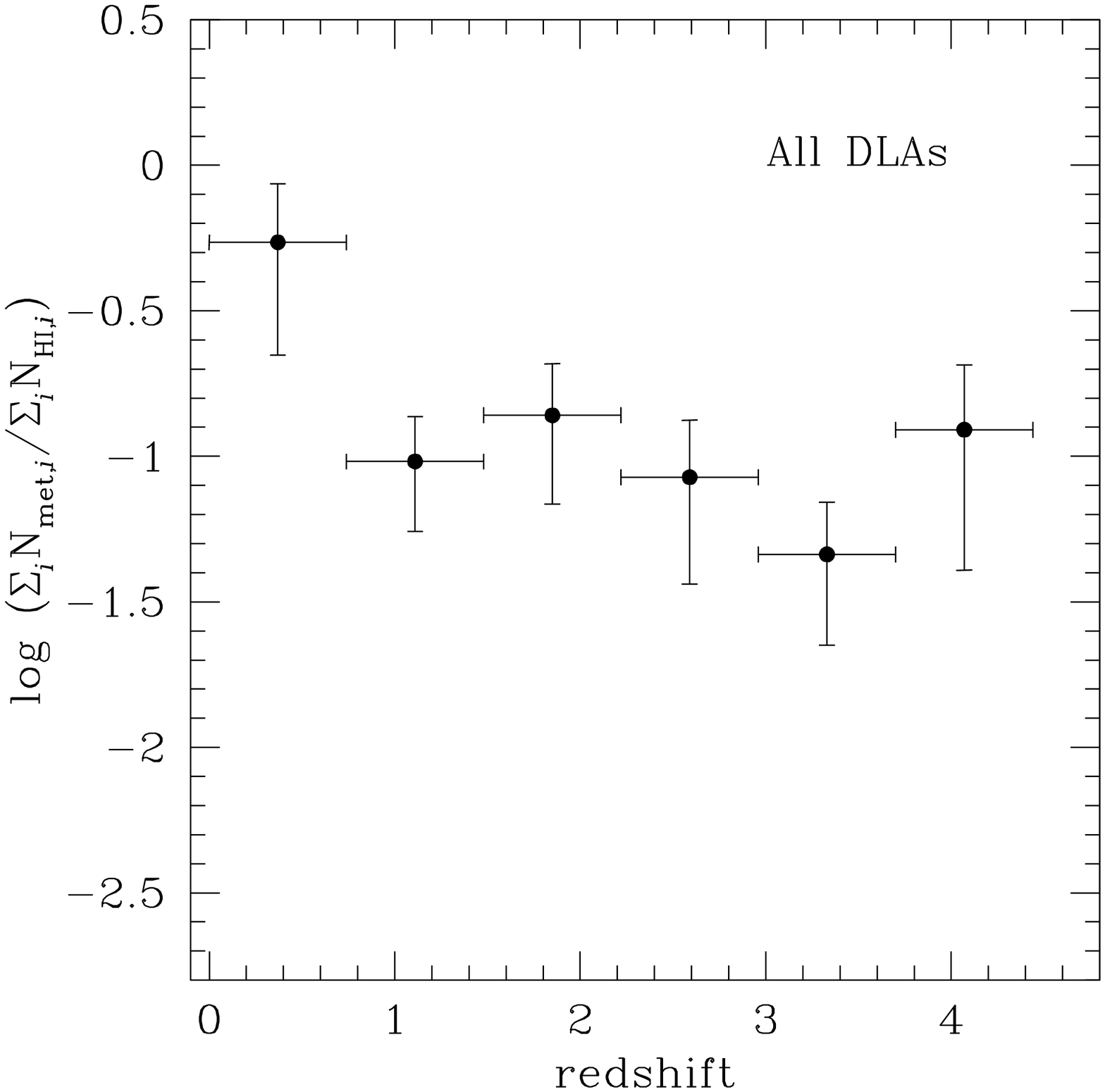}{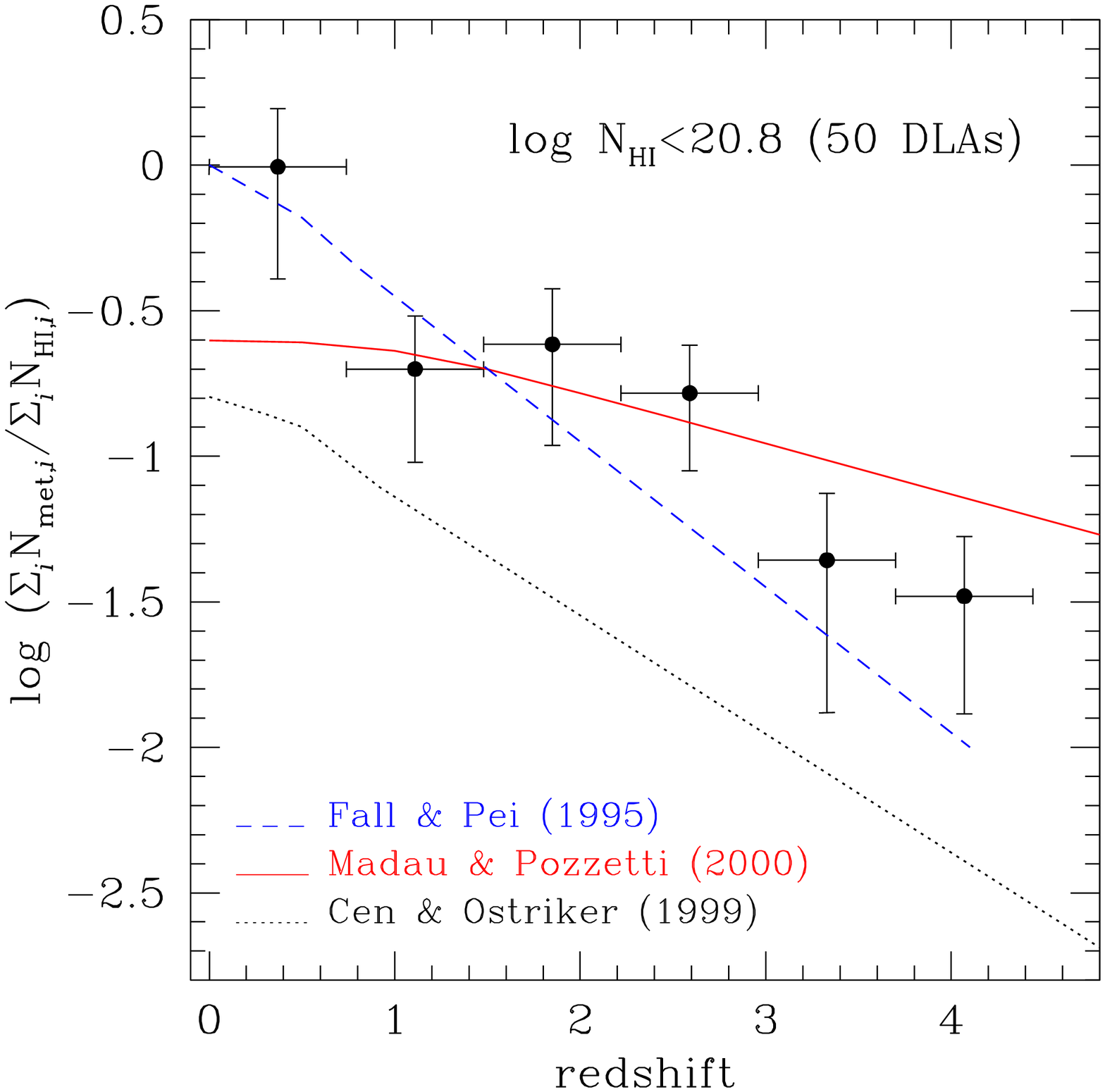}
\caption{Ratio of the total metal content to the total gas content as
a function of redshift. In the left panel the whole DLA sample is
used, whereas in the right panel the subsample of DLAs with $\log
N_{HI}<20.8$ is considered. Different curves in the right panel
represent the global chemical evolution of the Universe obtained by
different models.}
\end{figure}

The metallicity provided by DLAs is considered to be a good indicator
of the global chemical evolution of the Universe. In particular this
can be reproduced by the ratio of the total column density of metals
to the total column density of gas of DLAs, i.e.,
$Z\propto~\Sigma_i N_{met,i}/\Sigma_i N_{HI,i}$ (Pei \& Fall 1995;
Pei, Fall \& Hauser 1999). The ratio of the total metal content to
the total gas content of DLAs in redshift bins has been calculated
recently by Pettini et al.~(1999) using 40 ZnII measurements, but
these are actually 22 data points at $0.4 < z < 2.8$, if we exclude upper and
lower limits.  The so called ``column density--weighted'' metallicity
obtained by these authors does not show any redshift evolution. Basically the
same result has been obtained by Vladilo et al.~(2000) using a larger
ZnII database (27 ZnII measurements at $0.4 < z <3.4$) and by
Prochaska \& Wolfe (2000) who have explored this quantity at higher
redshifts using FeII instead (39 FeII measurements at $1.8<z<4.4$).
If this analysis is done using our larger sample (\nnhi~DLAs at
$0.0<z<4.4$) the results are not different (left panel of Fig.~9).

Considering the discussion of the previous section, this is
not surprising. DLAs with low and high HI column densities are not
distributed in the same way, in particular the latter subsample is
less homogeneous in the redshift range than the former. This means
that when we consider the total metals to total gas ratio, where the
high $N_{HI}$ DLAs have a much larger weight compared to low $N_{HI}$
DLAs, at low redshifts ($z<1.5$) the metallicity is underestimated
because high $N_{HI}$ DLAs are preferentially detected when dust
content, and therefore metallicity, is low (see Fig.~7). The detection of
high $N_{HI}$ DLAs is particularly hard in the redshift bin $z\sim1$
because the observational capabilities are at the limits for both HST and
ground based telescopes. This is evident from
Fig.~1, where the distribution of DLAs as a function of redshift has a
hole in the interval $0.8 < z <1.8$, also confirmed by the DLA
selection function (the combined redshift survey path lengths for
DLAs) reported by Mathlin et al.~(2000). At high redshifts (7 DLAs at
$z>3.7$), the weighted metallicity is basically dominated by a
single DLA at $z=4.224$ (Prochaska \& Wolfe, 2000) with large HI
column density ($\log N_{HI}\simeq20.8$) and large metallicity ([Fe/H]
$\simeq-0.35$, dust corrected). Indeed the column density weighted
metallicity goes down to $-1.5$ if we exclude this DLA from the
analysis.

The results change if we make a selection of DLAs using those which
are likely to be less affected by dust obscuration. The total metals
to total gas ratio in DLAs show evolution in the subsample of 50 DLAs
with HI column density lower than 20.8 (right panel of Fig.~9). This
goes from solar at $z\sim0.4$ to 1/30 solar at $z\sim4.1$. For
comparison we also plot the metal production in the Universe as given
by the most recent results on the global star formation history of the
Universe (solid line: Madau \& Pozzetti, 2000), and that obtained
using cosmological hydrodynamical simulations (dotted line: Cen \&
Ostriker 1999). The difference between these two behaviors is
remarkable, both in the normalization and in the slope.  The observed
evolution has been reproduced by Pei \& Fall (1995), who have used
closed--box, inflow and outflow models with results
similar to those used for the chemical evolution of the MW. These
models were constrained by the very limited observational results known
at that time on the DLA metallicity.

\section{Discussion}

It is not clear yet how well the mean HI weighted metallicity obtained
from current observations can represent the global metal enrichment of
the Universe because different observational biases might affect the
results.  As shown in Fig.~8, low and high $N_{HI}$ DLAs are
distributed in a different way (the latter are concentrated at the
center of the redshift interval while the former is much more
scattered). Moreover, the two correlations are consistent with each
other at the $\sim2\sigma$ level.  The lack of high $N_{HI}$ with high
metallicity, theoretically first considered by Pei and Fall (1995) and
then found by Boiss\'e et al. (1998), might correspond to the lack of
high $N_{HI}$ DLAs at $z<1.5$. If there were a metallicity evolution
also for high $N_{HI}$ DLAs, dust obscuration could be too large at
$z<1.5$ to make DLAs detectable (a DLA with $\log N_{HI}=21.0$ and
metallicity $\log Z_{DLA}/Z_\odot=-0.5$ would result in an extinction
of about 1 magnitude at Ly$\alpha$). Moreover, this is where UV
observations, required to detect Ly$\alpha$ absorption, are very
sensitive to any appreciable dust opacity.

In general, before using DLAs to explore the metallicity evolution in
the Universe, one should understand if and how DLA galaxies represent
the total galaxy population at different epochs. 
Obviously, this is not a trivial task because many factors have to be
taken into account some of which are very poorly known. When trying to
match Lyman break galaxies to DLA galaxies, one should consider that
the former are star forming galaxies for which the metallicity is
usually inferred (and not measured) from the stellar and gaseous
emissivity, while the latter are interstellar clouds with large HI column
densities in galaxies of undefined nature for which the metallicity is
measured from the absorption lines. On the other hand, when comparing
DLAs with the intergalactic medium, one has to consider that high
density regions like DLAs have very likely a different metallicity
history than the mean value in the Universe and/or in the diffuse
intergalactic medium. Cen \& Ostriker (1999) have pointed out that
higher density regions have higher star formation rates and after a
fast heavy--element enrichment, the metallicity evolves very slowly
compared to lower density regions, due to a suppression of the star
formation.
 
We like to stress that it is the combination of a large sample of DLAs
(\nnhi~objects), a wide redshift baseline ($0.0<z<4.4$) and an
accurate dust correction applied to all heavy elements measured, that
has allowed us to demonstrate that the metallicity of DLAs evolves
with redshift. Had we limited ourselves to using Zn only (detected in
31 DLAs) with no dust correction, we would have not detected the same
effect, even if ZnII lines are known not to suffer from strong
saturation and dust depletion. This is because even small corrections
are important when trying to detect elusive effects. In addition, Zn
is much less abundant than other elements such as Fe and Si and,
therefore, measuring its absorption lines is not always an easy task.
The combined use of as many elements as possible allows one to much
more efficiently track the metallicity evolution down to gas clouds that are very metal poor.

\acknowledgments I wish to thank the organizers of the IAU Symposium
in Manchester, in particular Martin Harwit and Mike Hauser. I am
grateful to Nino Panagia and Massimo Stiavelli for their important
contribution at an early stage of the project.

\newpage
\centerline{Discussion}
\vspace{5mm}

Charley Lineweaver: I am trying to understand what your $Z(z)$ results mean for the star formation rate.  Does the Fall \& Pei model you showed for $Z(z)$ (which fits your results) have a star formation rate which increases more steeply betwen redshifts 0 and 1 than most of the other rates we have seen over the last two days?

Sandra Savaglio:  According to Michael Fall it is not. In fact, the cosmic star formation rate history in the Pei \& Fall (1995) models is very similar to that inferred subsequently from the UV luminosity density history.  In particular, the SFR in the Pe-Fall models also rises by about a factor of 10 between $z=0$ and $z\sim 1$.

Richard Mushotzky: Most of the impact parameters are large, and so systematically sample outer, low-abundance regions of galaxies.

Michael Fall:  A random distribution of impact parameters, with more impacts at large galactocentric distances, is just what one needs to get the mass-weighted mean metallicity, which is the same as the column-density-weighted mean metallicity.  This is an unbiased measure of metallicity.

Savaglio:  I agree.

Eli Dwek:  What reference abundances did you use to correct for the depletions -- Solar or B-star abundances?

Savaglio:  I don't expect to find different results, globally speaking.

\end{document}